\documentclass[12pt]{article}
\usepackage{amssymb,amsmath,slashed,latexsym,xcolor}
\usepackage{float}
\allowdisplaybreaks
\newcommand{\ft}[2]{{\textstyle\frac{#1}{#2}}}

\def\rme{{\rm e}}
\def\rmi{{\rm i}}

\newcommand{\hc}{{\rm h.c.}}
\newcommand{\bbox}{\lower.2ex\hbox{$\Box$}}
\newsavebox{\uuunit}
\sbox{\uuunit}
    {\setlength{\unitlength}{0.825em}
     \begin{picture}(0.6,0.7)
        \thinlines
        \put(0,0){\line(1,0){0.5}}
        \put(0.15,0){\line(0,1){0.7}}
        \put(0.35,0){\line(0,1){0.8}}
       \multiput(0.3,0.8)(-0.04,-0.02){12}{\rule{0.5pt}{0.5pt}}
     \end {picture}}

\csname @addtoreset\endcsname{equation}{section}
\newcommand{\SU}{\mathop{\rm SU}}
\newcommand{\SO}{\mathop{\rm SO}}
\newcommand{\U}{\mathop{\rm {}U}}

\newcommand{\OSp}{\mathop{\rm {}OSp}}
\newcommand{\Symp}{\mathop{\rm {}Sp}}

   \usepackage[nosort]{cite}
   \pdfoutput=1
  \usepackage[pdftex]{hyperref}
\newcommand{\dr}{\raise.3ex\hbox{$\stackrel{\leftarrow}{\delta  }$}{}}
\newcommand{\dl}{\raise.3ex\hbox{$\stackrel{\rightarrow}{\delta }$}{} }
\newcommand{\pl}{\raise.3ex\hbox{$\stackrel{\rightarrow}{\partial }$}{} }

\begin{document}

\begin{titlepage}
\begin{flushright}
CERN-TH-2018-241
\end{flushright}
\vspace{.5cm}
\begin{center}
\baselineskip=16pt
{\LARGE Noether Supercurrents, Supergravity and  \\ \vskip 0.2cm  Broken Supersymmetry 
}\\
\vfill
{\large  {\bf Sergio Ferrara}$^{1,2,3}$ and {\bf Magnus Tournoy}$^4$} \\
\vfill

{\small$^1$ Theoretical Physics Department, CERN CH-1211 Geneva 23, Switzerland\\\smallskip
$^2$ INFN - Laboratori Nazionali di Frascati Via Enrico Fermi 40, I-00044 Frascati, Italy\\\smallskip
$^3$ Department of Physics and Astronomy and Mani L.Bhaumik Institute for Theoretical Physics, U.C.L.A., Los Angeles CA 90095-1547, USA\\\smallskip
$^4$   KU Leuven, Institute for Theoretical Physics, Celestijnenlaan 200D, B-3001 Leuven,
Belgium  \\[2mm] }
\end{center}
\vfill
\begin{center}
{\bf Abstract}
\end{center}
{\small
Some general aspects of supersymmetry and supergravity are briefly reviewed with emphasis on Noether supercurrents and their role in the discussion of supersymmetry breaking.
} \vfill

\begin{center}
{\it Contribution to the Proceedings of the Erice International School of Subnuclear Physics, \\56th Course: 
``From gravitational waves to QED, QFD and QCD"\\ Erice, 14-23 June 2018}
\end{center}

\hrule width 3.cm
{\footnotesize \noindent e-mails: Sergio.Ferrara@cern.ch,  magnus.tournoy@kuleuven.be}
\end{titlepage}

\addtocounter{page}{1}
 \tableofcontents{}
\newpage
\section{Introduction}

There are many phenomena in theoretical particle physics that await a proper explanation, even if one accepts the "dogmas" of Lagrangian renormalizable quantum field theory. Supersymmetry, a symmetry that enlarges the space-time symmetry of elementary particle interactions and their description in terms quantum field theory, is a long standing response to multiple questions that arose after the advent of the Standard model. Let us briefly recall some subjects that may have to do with supersymmetry \cite{Nath:2016qzm,Linde:2016bcz,Akrami:2018ylq,AlvarezGaume:1983gj,Weinberg:1988cp,Weinberg:2000yb,Akrami:2017cir}:
\begin{itemize}
\item Why two very distinct sets of fields which obey different statistics behave in a "democratic way" under the gravitational force?
\item Hierarchy problem (existence of fundamental scales of immensely different orders of magnitude) 
\item Estimates of proton lifetime in Grand Unified Theories 
\item Why the top-quark mass is comparable to the Higgs mass but much larger
than other quark masses in the Standard Model?
\item If “inflation” is the correct dynamics of the Early Universe, who is the
inflaton, the scalar degree of freedom that drives it?
\item Where do “dark matter” and “dark energy” come from?
\item The cosmological constant problem.
\end{itemize}
In this brief report we review the basics of supersymmetry and supergravity, their Noether currents and spontaneous supersymmetry breaking in rigid and local supersymmetry. Note that this contribution is not intended for experts but for a mixed audience of mainly PhD students in high energy physics.
\section{Noether currents and supercurrents}
We start by discussing Noether's theorem in Lagrangian field theory and will focus specifically on space-time symmetries. \\
Let us consider (Super) Lie group symmetries of a local Lagrangian field theory with Lagrangian ${\cal L}$. In the infinitesimal limit the corresponding (Super) Lie algebra generators $Q_A$ must give rise to a vanishing variation\footnote{We use here the action of symmetry operators after quantization. In the classical canonical formalism, symmetry transformation are generated by Poisson brackets
\begin{equation}
\delta{\cal L}=\left\{\epsilon^A Q_A,{\cal L}\right\}_{\text{PB}}\,.
\end{equation}}\textsuperscript{,}
\footnote{We adopt the conventions of \cite{Freedman:2012zz} unless for superspace formulations where \cite{Wess:1992cp} is followed.}
\begin{equation}
\delta {\cal S}=\int d^4 x\, \delta {\cal L}=0\,,\qquad \delta{\cal L} =-\rmi\left[\epsilon^A Q_A, {\cal L}\right]\,,
\end{equation}
where the index $A$ runs over the different symmetries parametrised by the constants $\epsilon^A$. In superalgebras some of the $\epsilon^A$ are anticommuting parameters and consequently their generators $Q_A$ will satisfy anticommutation relations.

Noether's theorem asserts that, in the situation described above, for every symmetry a vector current $J_{\mu A}(x)$ must exist which is conserved as a consequence of the field equations. In the case of space-time (super)symmetries the Lagrangian transforms as a total derivative
\begin{equation}
\delta{\cal L}=\epsilon^A\partial_\mu K^\mu_A\,.\label{varL}
\end{equation}
The Noether current then takes the form 
\begin{equation}
\epsilon^A J^\mu_A=-\delta\phi^i\frac{\pl{\cal L}}{\partial \partial_\mu \phi^i}+\epsilon^A K^\mu_A\,. \label{Ncur}
\end{equation}
The indices on the fields should be thought of as denoting the specific representation of the fields in the algebra.
The current conservation is readily seen from the field equations\footnote{We reserve $\approx $ for identities that are valid due to field equations.}
\begin{equation}
\frac{\dl {\cal S}}{\delta \phi ^i }=\frac{\pl {\cal L}}{\partial\phi^i}-\partial_\mu\frac{\pl {\cal L}}{\partial \partial_\mu\phi^i}\approx 0\,.
\end{equation}
Taking the total derivative of the Noether current in \eqref{Ncur} while contracting the coordinate indices, we find by plugging in the field equations
\begin{align}
\epsilon^A\partial_\mu J^\mu_A&=-\partial_\mu\left(\delta\phi^i\frac{\pl{\cal L}}{\partial \partial_\mu \phi^i}\right)+\epsilon^A \partial_\mu K^\mu_A\nonumber\\
&\approx-\partial_\mu\delta\phi^i\frac{\pl{\cal L}}{\partial \partial_\mu \phi^i}-\delta\phi^i\frac{\pl {\cal L}}{\partial\phi^i}+\epsilon^A \partial_\mu K^\mu_A\,.
\end{align}
The first two terms on the last line are simply the symmetry variation of the Lagrangian and therefore, because of \eqref{varL} they should cancel with last term. The conservation of the current is thus ensured by the field equations
\begin{equation}
\partial_\mu J^\mu_A\approx 0\,.
\end{equation}
The symmetry variation of the fields contain an intrinsic part dependent on their specific representation and an orbital part that is caused by the action of the symmetry operator on the coordinates
\begin{align}
\delta \phi^i= \hat\delta \phi^i-\delta x^\sigma\partial_\sigma \phi^i&\,,\quad \delta x^\sigma = \epsilon ^A \Delta_A^\sigma\,,\quad \hat\delta\phi^i=\epsilon^A\theta_A^i(x)\nonumber\\
\epsilon^A K^\mu_A=&-\delta x^\mu{\cal L}=-\epsilon^A\Delta^\mu_A {\cal L}\,. \label{symvar}
\end{align}
For linear symmetries we have $\theta^i_A=T_{A}^{ i }{}_j\phi^j$. The last line is a simple consequence of the fact that only the orbital spacetime transformations cause the Lagrangian to transform as a total derivative. This fact together with the form of the transformation of the fields in \eqref{symvar} inform us that the (bosonic) spacetime currents share a similar structure. By using that the currents, obtained by the Noether procedure, are not unique but can be modified by adding a term
\begin{equation}
J_\mu^A \,\rightarrow \,J_{\mu}^A+\partial^\nu S^A_{\nu\mu}\quad \text{ with }\quad S^A_{\nu\mu}=-S^A_{\mu\nu}
\end{equation}
it is possible to add improvement terms to the canonical energy-momentum tensor, which is the Noether current of spacetime translations,
\begin{equation}
J_{\mu A}(A=\nu)\,\rightarrow\,\theta_{\mu\nu}=J_{\mu\nu}+\left(\partial_\mu\partial_\nu-g_{\mu\nu}\Box\right)f
\end{equation}
in such a way that all Noether currents of the bosonic spacetime symmetry group can be expanded as "moments" of $\theta_{\mu\nu}(x)$ 
\begin{equation}
J_{\mu A}\epsilon^A=\theta_{\mu\nu}\delta x^\nu\quad \text{so that}\quad J_{\mu A}=\theta_{\mu\nu}\Delta^\nu_A\,.
\end{equation}
For the conformal algebra, the maximal spacetime symmetry, the coordinate variations $\delta x^\nu$ satisfy the differential equation
\begin{equation}
\partial_\mu \delta x_\nu +\partial_\nu \delta x_\mu=\frac2D \eta_{\mu\nu}\partial^\rho\delta x_\rho\quad\rightarrow\quad \SO(D,2) \text{ for } D>2 \,.
\end{equation}
The solution of the equation is 
\begin{equation}
\delta x^\mu = a^\mu +\lambda^{\mu\nu} x_\nu+\lambda_D x^\mu +\left(x^2\lambda_K^\mu-2x^\mu x\cdot \lambda_K\right)\,.
\end{equation}
From this equation we can find the charges of the Poincar\'e and conformal algebra, which are the momenta of $\theta_{0\mu}$,
\begin{equation}
Q_A=\int d^3 x\, J_{0A}=\int d^3 x \,\theta_{0\nu}\Delta_A^\nu \quad \Rightarrow \quad 
\begin{cases}
P_\mu=\int d^3x\, \theta_{0\mu}\\
 M_{\mu\nu}=\ft12\int d^3 x\, \left(x_\nu\theta_{0\mu}-x_\mu\theta_{0\nu}\right)\\
 D=\int d^3 x\, \theta_{0\mu} x^\mu\\
  K_\mu=\int d^3 x\,\left(x^2\theta_{0\mu}-2 x_\mu x^\sigma\theta_{0\sigma}\right)
\end{cases}\,.
\end{equation}
For supersymmetry the vector-spinor current $J_{\mu A}(A=\alpha)$ can be improved
\begin{equation}
J_{\mu A}(A=\alpha)\,\rightarrow J_{\mu\alpha}^{\text{impr}}=J_{\mu\alpha}+(\gamma_{\mu\nu}\partial^\nu\psi)_\alpha
\end{equation} 
such that the general (superconformal) supercurrent  can be written as $\epsilon^\alpha J_{\mu \alpha}^{\text{impr}}$, with $\epsilon=\epsilon_0+\slashed{x}\epsilon_1$. Both the improved stress tensor and the improved fermionic supercurrent, are constrained by the requirement of current conservation
\begin{alignat}{2}
&\text{Poincar\'e symmetry:} \qquad \qquad\qquad &&\partial^\mu\theta_{\mu\nu}\approx 0\,,\quad \theta_{\mu\nu}\approx\theta_{\nu\mu}\\
& \text{Super-Poincar\'e symmetry:} &&\partial^\mu J_{\mu\alpha}^{\text{impr}}\approx 0\\
&\text{Conformal symmetry:} &&\theta^\mu{}_\mu \approx 0\\
&\text{Superconformal symmetry:	} &&(\gamma^\mu)_\alpha{}^\beta J_{\mu\beta}^{\text{impr}} \approx 0\,.
\end{alignat}
The spinor charges are obtained from the space integrals of the time components of the corresponding conserved currents
\begin{equation}
Q_\alpha=\int d^3 x\, J_{0\alpha}^{\text{impr}}\,,\quad S_\alpha=\int d^3 x\, \left(\gamma^\mu x_\mu\right)_\alpha{}^\beta J^{\text{impr}}_{0\beta}\,.
\end{equation}
For superconformal symmetry we also need the $R$-symmetry current $J_{\mu 5}$. Altogether the currents combine into a multiplet
\begin{equation}
(J_{\mu 5},J^{\text{impr}}_{\mu\alpha},\theta_{\mu\nu})\,.
\end{equation}
For Poincar\'e (broken conformal) supersymmetry the otherwise vanishing parts of the current multiplet now form a chiral multiplet on their own \footnote{For chiral superfields $z^i$ with superpotential $W(z^i)$ then $A\varpropto W-\frac13 z^iW_i$.}
\begin{equation}
S=\left(A,\psi,F\right)\qquad \left(\psi=\gamma^\mu J^{\text{impr}}_{\mu}, F=\theta_\mu{}^\mu+\rmi\partial^\mu J_{\mu 5}\right)\,. \label{ccurrent}
\end{equation}
\section{Navigating in superspace}
Supersymmetry's most natural habitat is superspace. In superspace the spacetime is extended by Grassmann coordinates. In the case of ${\cal N}=1$, $D=4$ the supermanifold ${\cal M}_{4|4}$ will be described by the coordinates $(x^\mu,\theta_\alpha)$ with $\theta_\alpha\theta_\beta=-\theta_\beta\theta_\alpha$. The superalgebra acts on $(x^\mu,\theta_\alpha)$ as a shift\cite{Ferrara:1974ac}
\begin{equation}
\theta_\alpha\,\rightarrow\,\theta_\alpha+\epsilon_\alpha\,,\quad x^\mu\,\rightarrow\,x^\mu+\rmi\left(\theta\sigma^\mu\bar\epsilon-\epsilon\sigma^\mu\bar\theta\right)\,.
\end{equation}
For a superfield $\Phi(x,\theta)$ the supersymmetry variation is
\begin{equation}
\delta\Phi=\epsilon Q\,\Phi=\left[\epsilon\frac{\partial}{\partial \theta}+\bar\epsilon\frac{\partial}{\partial\bar\theta}+\rmi\left(\theta\sigma^\mu\bar\epsilon-\epsilon\sigma^\mu\bar\theta\right)\partial_\mu\right]\Phi\,.
\end{equation}
We can covariantize the spinor derivatives
\begin{equation}
	D_\alpha=\frac{\partial}{\partial\theta^\alpha}+\rmi\left(\sigma^\mu\bar\theta\right)_\alpha\partial_\mu\,,\quad \overline{D}_{\dot\alpha}=-\frac{\partial}{\partial \bar\theta^{\dot\alpha}}-\rmi\left(\bar\theta\sigma^\mu\right)_{\dot\alpha}\partial_\mu\,,\quad \left\{D_\alpha,\overline D_{\dot\alpha}\right\}=-2\rmi\sigma_{\alpha\dot\alpha}^\mu\partial_\mu
\end{equation}
such that they anticommute with the $Q_\alpha$ operators
\begin{equation}
\left\{D_\alpha,Q_\beta\right\}=\left\{D_\alpha,\overline{Q}_{\dot\beta}\right\}=0\,.
\end{equation}
Therefore covariant derivatives of superfields are again representations of the superalgebra. The supercurrent multiplet obeys the following structure
\begin{equation}
\overline{D}^{\dot\alpha}J_{\alpha\dot\alpha}\equiv\overline D^{\dot\alpha} \left(\sigma_{\alpha\dot\alpha}^\mu J_\mu\right)=D_\alpha S\,, \label{ccons}
\end{equation}
where $S$ is a chiral multiplet $\overline D_{\dot\alpha} S=0$. The superspace expansion of supercurrent multiplet is \cite{Ferrara:1974pz}
\begin{align}
 J_\mu(x,\theta,\bar\theta)=&\,J_\mu^5(x)+\left(\frac\rmi{4}\theta^\alpha \left[-3J^{\text{impr}}_{\mu\alpha}(x)+\left(\sigma_\mu\overline\sigma J^{\text{impr}}\right)_\alpha\right]+2\rmi\theta^2\partial_\mu\left( A+\rmi  B\right)+\hc\right)\nonumber\\
 &-\theta\sigma^\rho\bar\theta\left(\frac12\left[-3\theta_{\rho\mu}+\eta_{\rho\mu}\theta_\lambda{}^\lambda\right]-\frac14\epsilon_{\rho\mu\nu\lambda}\partial^\nu J^{5\lambda}\right)+\ldots
\end{align}
The chiral multiplet we mentioned in \eqref{ccurrent} is encapsulated inside the current multiplet and whenever the symmetries are extended to the full superconformal algebra it vanishes individually. The Ward identity 
\begin{equation}
-\frac18(\gamma_{(\lambda})^{\alpha\dot\beta} \left\{ J^{\text{impr}}_{\mu)\alpha},\overline Q_{\dot\beta}\right\}_{\text{PB}} = \theta_{\lambda\mu}
\end{equation}
implies that when supersymmetry is unbroken $Q_\alpha|0\rangle=0$ the vacuum energy vanishes $\langle{\theta^{\lambda\mu}}\rangle=0$ \cite{Zumino:1974bg}.
\section{Supercurvatures of local supersymmetry}
In general relativity the coordinate covariant derivatives don't commute
\begin{equation}
\left[\nabla_\mu,\nabla_\nu\right]V^\rho=R_{\mu\nu}{}^\rho{}_\sigma V^\sigma-T_{\mu\nu}{}^\sigma\nabla_\sigma V^\rho\,.
\end{equation}
The first term defines the curvature tensor and the second the torsion tensor. In the presence of torsion, the Palatini (first-order) formalism is not equivalent to the ordinary Riemannian formulation. For example, a Majorana fermion couples to the torsion via the spin connection and a new term $k^2(\psi\psi)$ is present in the final theory.\\
In superspace, the supertorsion and supercurvature satisfy some constrained equations, so that the geometry is finally encoded in three superfields \cite{Ferrara:1977mv,Wess:1977fn,Wess:1978bu,Wess:1978ns}
\begin{alignat}{2}
&W_{\alpha\beta\gamma}(x,\theta)\quad &&\rightarrow\quad \text{Weyl tensor}\,,\\
&E_\mu(x,\theta,\bar\theta)\quad &&\rightarrow\quad \text{Einstein tensor}\\
&{\cal R}(x,\theta)\quad &&\rightarrow\quad \text{Scalar curvature}
\end{alignat}
The superfields are subject to certain constraints. For example \cite{Ferrara:1977mv}:\footnote{The Ward identity relating the Weyl tensor with the Einstein tensor multiplet is 
\begin{equation}
D^\alpha W_{\alpha\beta\gamma}=\frac12\rmi\left(\partial_{\beta}{}^{\dot\gamma} E_{\gamma\dot\gamma}+\partial_{\gamma}{}^{\dot\gamma}E_{\beta\dot\gamma}\right).
\end{equation}}
\begin{equation}
\overline D^{\dot\alpha}E_{\alpha\dot\alpha}=D_\alpha{\cal R}\,.\label{ERrel}
\end{equation}
which implies
\begin{equation}
\nabla^\mu G_{\mu\nu}=0\,,\quad G_\mu{}^\mu=-R\,.
\end{equation}
The equivalence of the Ward identity in \eqref{ERrel} with the structure of the supercurrent in \eqref{ccons} is not an accident. Since that the Einstein tensor is part of the $E_\mu|_{\theta\sigma^\nu\overline\theta}$ component, the graviton field equation together with supersymmetry tells us that
\begin{equation}
2 E_{\alpha\dot\alpha}\approx \kappa J_{\alpha\dot\alpha}\,.
\end{equation}
The action of ordinary linearised supergravity can thus be written as \cite{Ferrara:1977mv}
\begin{equation}
\int d^4 x d^4\theta\left(V^{\alpha\dot\beta}E_{\alpha\dot\beta}-\frac{\kappa}{2}V^{\alpha\dot\beta}J_{\alpha\dot\beta}\right)\,,
\end{equation}
with $V_{\mu}|_{\theta\sigma\overline\theta}=h_{\mu\nu}-\eta_{\mu\nu}h$. The multiplet $E_{\alpha\dot\alpha}$ is also called the Ferrara-Zumino multiplet \cite{Ferrara:1977mv}. The full component structure of this multiplet together with the ones of the other supercurvature multiplets were derived in \cite{Ferrara:1988qx}. Recently the correspondance between field equations and currents has been applied to construct the supercurrent and curvature multiplets in ${\cal N}=1$ conformal supergravity \cite{Ferrara:2017yhz,Ferrara:2018dyt}. The Ward identities are then relations between field equations. Because of the extensiveness of the superconformal algebra the cases for different ${\cal N}=1$ Poinar\'e supergravities (old minimal, new minimal, $16+16$) are all deducible from the conformal results. Different realtizations of supercurrents and curvature multiplets were also studied in \cite{Kuzenko:2010am,Kuzenko:2012vd}\\
In contrast with Riemannian geometry, "flat" superspace has non-vanishing (super)torsion 
\begin{equation}
T^{C}_{AB}\quad\rightarrow\quad T^c_{\alpha\dot\beta}=2\sigma^c_{\alpha\dot\beta}\,.
\end{equation}
Hence the superspace geometry underlying supergravity is not (super)Riemannian geometry \cite{Wess:1977fn,Wess:1978bu,Wess:1978ns,Arnowitt:1975xg,Sohnius:1982xs,Ferrara:1978em}.
\section{Spontaneous supersymmetry breaking}
In the case spontaneous broken supersymmetry the Noether currents are conserved but the vacuum is not invariant. The solutions of the field equations do not preserve the symmetry
\begin{equation}
\delta\phi^i\,\rightarrow\,\langle\delta\phi^i\rangle\neq 0\qquad \langle\left[\epsilon^A Q_A,\phi^i\right]\rangle\neq 0\,\Rightarrow\, Q_A|0\rangle\neq 0\,.\label{varf}
\end{equation}
The Goldstone theorem tells us that there is a massless field lying in the direction of the broken symmetry. For superfields \eqref{varf} implies that higher components must have vanishing VEV. Hence, looking at higher components of superfields leads to a model independent definition of supersymmetry breaking. A superfield expanded in fermionic coordinates
\begin{equation}
\phi(x,\theta)=\sum\theta^{\alpha_1}\ldots\theta^{\alpha_n}\phi_{\alpha_1\ldots\alpha_n}(x)
\end{equation}
transforms under supersymmetry as
\begin{equation}
\delta\phi(x,\theta)=-\rmi\left[\epsilon Q,\phi(x,\theta)\right]\,.
\end{equation}
This translates for the components of the superfield into
\begin{equation}
\delta\phi_i=\epsilon\partial\phi_{i-1}+\epsilon\phi_{i+1}
\end{equation}
and therefore for the last and first component
\begin{equation}
\delta\phi_{\text{last}}=\epsilon\partial \phi_{\text{last}-1}\,,\quad \delta\phi_0=\epsilon\phi_1\,.
\end{equation}
With broken supersymmetry $\langle \phi_1\rangle\neq 0$, and more generally
\begin{equation}
\langle \delta\phi_i\rangle=\epsilon\langle\partial\phi_{i-1}+\phi_{i+1}\rangle\neq 0\,.
\end{equation}
Some examples are the chiral scalar and gaugino multiplet
\begin{alignat}{2}
S&=A+\sqrt2 \theta \psi+ F\qquad &&\langle F\rangle\neq 0\\
W_\alpha&=-\rmi\lambda_\alpha+\left(\delta_\alpha^\beta D -\rmi\sigma^{mn}{}_{\alpha}{}^\beta F_{mn}\right)\theta_\beta+\theta^2\sigma_{\alpha\dot\alpha}^m\partial_m\overline\lambda^{\dot\alpha}\qquad &&\langle D\rangle\neq 0\,.
\end{alignat}
Notice that a vacuum value of the lowest component does not break supersymmetry, since the first component is not a next "field".\footnote{We assume that the superfields are primary in the sense that they cannot be written as covariant derivatives of other superfields.} If we apply this reasoning to the supercurvature multiplets, we can classify "curved" supersymmetric backgrounds other than Minkowski spacetime \cite{Ferrara:2017yhz,Ferrara:2018dyt}
\begin{equation}
W_{\alpha\beta\gamma}|_{\theta}=0 \rightarrow "\text{Weyl tensor}"=\partial_{[\mu}E_{\nu]}|_{\theta=0}=0 \,,\quad E_{\alpha\dot\alpha}|_{\theta=0}\neq 0\,,\quad {\cal R}|_{\theta=0}\neq 0\,.
\end{equation}
This approach is different from the one followed in \cite{Festuccia:2011ws,Dumitrescu:2012ha} where, to derive possible supersymmetry preserving backgrounds, the requirement of vanishing supersymmetry variation of the gravitino field was used.
\section{Backgrounds with four preserved supersymmetries}
The backgrounds that preserve four supersymmetries, and are
not Minkowski spacetime, are the ones where the first component of the scalar curvature
is non-vanishing or the first component of the Einstein tensor is non-vanishing
\begin{align}
&{\cal R}|_{\theta=0}=\bar u\neq 0\qquad (\text{AdS}_4)\\
&E_{\alpha\dot\alpha}|_{\theta=0}=A_{\alpha\dot\alpha}\neq0\qquad \left(S_3\times L,\,\text{AdS}_3 \times L\right)\,.
\end{align}
$u$ and $A_{\alpha\dot\alpha}$ are the (six bosonic) auxiliary fields of old minimal supergravity, with the action
\begin{equation}
\rme^{-1}{\cal L}=\frac1{2\kappa^2}\left[R-\overline{\psi}_\mu\gamma^{\mu\nu\rho}{\cal D}_\nu\psi_\rho-6u\overline u + 6A_\mu^2\right]\,.
\end{equation}
All of the supersymmetry preserving backgrounds require that the Weyl tensor vanishes $W_{\alpha\beta\gamma}=0$. The superalgebras of the backgrounds are all subalgebras of the superconformal algebra 
\begin{align}
&\OSp (1|4)\supset\Symp (4,\mathbb{R})\sim \SO (3,2)\\
&\SU (2,1)\times\SU(2) \supset \SU (2) \times \SU( 2)\times \U (1)=\SO (4)\times \U (1)\\
&\SU (2,2)\times\SU (1,1)\supset \SU (1,1)\times\SU (1,1)\times \U (1)=\SO (2,2)\times \U (1)\,.
\end{align}
The necessary and sufficient conditions for unbroken supersymmetry can be derived by requiring that the supersymmetry variations of the components of the curvature multiplets vanish. For example from the variation of the $\theta$-component of the Einstein tensor multiplet one finds that the vector component of the Einstein multiplet, which contains the Einstein tensor, must equal zero \cite{Wess:1978ns,Sohnius:1982xs,Ferrara:1978em,Stelle:1978ye,Sohnius:1981tp,Breitenlohner:1976nv}  
\begin{align}
E^{\text {symm}}_{\mu}|_{\theta\sigma^\nu\overline\theta}&=R_{\mu\nu}-2A_\mu A_\nu+2\eta_{\mu\nu}A^\rho A_\rho+3\eta_{\mu\nu}u\overline u=0\,.
\end{align}
The other conditions for preserved supersymmetry in old minimal supergravity backgrounds are \cite{Ferrara:2017yhz,Ferrara:2018dyt,Festuccia:2011ws,Dumitrescu:2012ha}
\begin{equation}
W_{\alpha\beta\gamma\delta}=0\,,\quad u A_\mu=0\,,\quad \nabla_\mu A_\nu=0\,,\quad \partial_\mu u=0\,.
\end{equation}
From the constraints imposed by the curvature multiplets we derive the following backgrounds
\begin{align}
\text{AdS}_4&\,\rightarrow\,A_\mu=0\quad u\neq 0\quad R+12 u\overline u=0\\
{\cal M}_3\times L&\,\rightarrow\, A_\mu\neq 0\quad u=0\quad R+6A^\mu A_\mu=0\nonumber\\
&\qquad A_0\neq 0\,\Leftrightarrow\, S_3\times L\text{ and } A_3\neq 0\,\Leftrightarrow\, \text{AdS}_3\times L\,.
\end{align}
The other two backgrounds are flat space with $u= A_\mu =0,\, R_{\mu\nu}=0$ and a Nappi-Witten background with $A_\mu$ lightlike and $u=0$ ($R=0$, but $R_{\mu\nu}$ is not zero since $R_{\mu\nu}=2A_\mu A_\nu$)\cite{Nappi:1993ie,Figueroa-OFarrill:1999cmq}.
\section*{Acknowledgments}
We would like to thank A.Kehagias for discussions. We thank A. van Proeyen and M. Samsonyan for collaboration on the topics of this review.  
The work of S.F. is supported in part by CERN TH Department and INFN-CSN4-GSS. The work of M.T. is supported in by the KU Leuven C1 grant ZKD1118
C16/16/005 and the FWO odysseus grant G.0.E52.14N.

\providecommand{\href}[2]{#2}\begingroup\raggedright\endgroup

\end{document}